\documentstyle{article}

\begin{document}

\begin{center}
{\huge\bf On Quantum Cohomology}
\end{center}

\vspace{1cm}
\begin{center}
{\large\bf
F.GHABOUSSI}\\
\end{center}

\begin{center}
\begin{minipage}{8cm}
Department of Physics, University of Konstanz\\
P.O. Box 5560, D 78434 Konstanz, Germany\\
E-mail: ghabousi@kaluza.physik.uni-konstanz.de
\end{minipage}
\end{center}

\vspace{1cm}

\begin{center}
{\large{\bf Abstract}}
We discuss a general quantum theoretical example of quantum  
cohomology and show that various mathematical aspects of quantum  
cohomology have quantum mechanical and also observable significance.
\end{center}

\begin{center}
\begin{minipage}{12cm}

\end{minipage}
\end{center}

\newpage

The quantum cohomology is one of the most fundamental and  
intressting mathematical-physical fields and although it is  
introduced according to certain physical models \cite{Witva},  
however it should be considered as a general invariant geometrical  
tool for all quantum theories. Nevertheless, in view of various  
mathematical dificulties \cite{sadov} its physical foundations are  
not well discussed yet. The main reason for this situations lies  on  
the non-well understood toplogical or differential geometric  
structure of quantization as a general sheme.

It is importent to mention that if one take the fact serious that  
classical mechanics is a classical limit of quantum mechanics, then  
a fundamental part of topology which is based on the globalization  
of classical mechanical results, e. g. Morse theory and symplectic  
topology, should be considered as a classical limit of some quantum  
topological originals. In view of the fact that the main difference  
between quantum and classical mechanics is the global (topological)  
character of states and accordingly the observables of quantum  
mechanics despite of local character of classical observables  
\cite{vier};
It is natural that the main difference in the classical and quantum  
geometries also arise in the topological scope. In other words, in  
view of the genuin topological character of quantization it is quite  
natural that quantization has such an influence like a quantum  
deformation of cohomology on the topology of the quantized system.

Moreover, in view of the necessary symplectic background of  
quantization it is also not surprising that the quantum cohomology  
becomes equivalent to some generalization of certain results on  
invariant structures of symplectic mechanics (in quantum theoretical  
sense), i. e. to the so called Floer cohomology \cite{Nash}.

Briefly speaking the quantum cohomology should be considered as a  
result of existence of flat connections, which are related with  
quantization, together with the multiply connectedness of the  
quantum phase space which is related with multivalued functions.  
Equivalently, a closed "path" (circle) surrounding the minimum cell  
of the quantized phase space with an area $h>0$ can not be shrunk to  
a point. It can be considered also as a result of finiteness of  
some relevant measures like position, i. e. position uncertainty  
$\delta q$ which are prevented to become zero in quantum mechanics  
$(\delta q > 0)$. In this sence, for example, the usual notion of "  
path" or {\it boundary} loses its definition in quantum phase space  
and we have a deformation of the original classical homologies and  
cohomologies, whereas the usual "classical" homologies and  
cohomologies should be considered as valid only in the classical  
phase space which is a simply connected manifold. To be precise, let  
us mention that in view of the uncertainty relation $\delta q  
\delta p = \hbar> 0$ a ring of width $\delta q$ in quantum phase  
space can not be shrunk to a circle, i. e. to the classical boundary  
of a $(2-D)$ manifold, whereas this is possible in a clacssical  
phase space where $\hbar = 0$. Considering the finiteness of the  
most minimal energy in quantum mechanics of a harmonic oscillator  
$(\delta E = 2 E_0 > 0)$ one should also recall the structure of  
original Morse function as the main ingredient of invariant  
structures on manifolds which is an energy functional. Hence, it is  
the ground state energy which differs the classical cohomology of  
harmonics, $\Delta Harm^0 = 0$, from the quantum cohomology of  
ground state, i. e. the Floer cohomology.

Coming back to the question of multiply connectedness it seems also  
natural to use the so called Novikov ring \cite{sadov} of  
multivalued functions to construct the quantum cohomology: Because  
on the one hand we have the natural relation between the multiply  
connectedness of a manifold and the multivalued functions and on the  
other hand we know about the quantization of angular momentum which  
results from a transformation of the related multivalued function  
into a single valued one \cite{dreh}. Thus, the general role of  
multivalued functions in quantum cohomology should be understood if  
one recalls that they represent the wave functions before their  
quantization.

To make furhter relations of quantum cohomology with physical  
observable effects transparent let us mention that our intrest on  
this field arose from
the question of edge currents \cite{kk} and potential drops  
\cite{dk} in quantum Hall effects (QHE) and also from the  
equivalence between the quantzation of Hall measures, i. e.  
resistivity and conductivity, and the flux quantization in  
superconductivity \cite{meinq}. Since, the question of boundary  
which is the main ingredient of homology plays an essential role in  
these phenomena, i. e. in the edge current, potential drops and flux  
quantization.
Furthermore, both QHE and the cohomology ring of a $2-D$ manifolds  
are topologically invariant structures, thus it is relevant to look  
for some relation between them. However, the first one is a quantum  
structure, whereas
the second one is a classical structure, thus one should ask if  
there is a quantum version of cohomology which should be related to  
the topological QHE structure? Therefore, the questions of boundary  
and also the relative cohomology can be helpfull here. Specially,  
the absense of the notion of "{\it path}" in quantum theory which  
can be considered in the closed case as the boundary of a $(2-D)$  
manifold, gives the right instrument to define non-trivial  
cohomologies.

\medskip
We will discuss here the general quantum mechanical foundations of  
quantum cohomology and show its equivalence with the Floer  
cohomology. In a subsequent paper \cite{next} we discuss also the  
mathematical questions which arise in proving this equivalence for  
the original models \cite{witva} and we will clarify the background  
structure of quantum cohomology of these models according to the  
discussion in Ref. \cite{sadov}.

Although, the structure of mentioned models of quantum cohomology  
is (in view of their various "extra" degrees of freedom) rather  
complicated to show the sufficiency of the pure quantization sheme  
of these models for the emergence of quantum cohomology directly in  
these cases. However, we give at least some intuitive reasons for  
our statement on the fundamentality and generality of the appearence  
of quantum cohomology in all {\it quantized} theories which can be  
proved.

First, recall that at least in the $(2+1)$ dimensional case the  
supersymmetry should be considered as a result of Poincare duality  
of $H^2 \cong H^0$ of the quantized theory. In other words, a  
supersymmetry requirement in suitable degrees of freedom manifests  
such a Poincare duality which is essential for a consistent quantum  
structure and also for a quantum cohomology. Therefore, the reason  
why such supersymmetric models \cite{witva} demonstrate quantum  
cohomologies should lies in their genuin (constructed) Poincare  
duality which is related with their geometric quantization.

Secondly, in view of the fact that every quantization sheme has to  
be equivalent to the canonical quantization, we have a classical  
symplectic background for all theories which are ${\it quantized}$.  
On the one hand, we know from the geometric quantization \cite{wood}  
that quantization of a classical phase space can be given according  
to the structure of the (complex) line bundle over the phase space  
which is equivalent to a principal $U(1)$-bundle over the same. On  
the other hand, a quantization of a symplectic structure of the  
classical phase space is equivalent to its complexification at least  
in the Heisenberg approach which is  equivalent to the stablishing  
of a Kaehler structure on the classical phase space. Thus, we have  
by the quantization of a theory a transition from the classical  
cohomology as the invariant structure of its classical phase space  
to an other invariant structure of its quantized phase space. It is  
this invariant structure which should be considered as the quantum  
cohomology for the given theory, hence its classical limit becomes  
the mentioned classical cohomology. In other words, the quantum  
cohomology is a topologically  invariant result of the quantum  
deformation on a given phase space.

To show these circumstances in mentioned models, let us mention  
that for example in the first model the Kaehler structure and a $H^1  
=0$ requirement  are used to obtain the "invariants" of related  
quantum cohomology. Now $H^1 = 0$ means that one is considering {\it  
only} flat connections or properly the moduli space of instantons  
of the model. Flat connections in turn means ( in the proper $U(1)$  
sense ) that we have to do with  the quantization sheme according to  
the flat $U(1)$-connection.

Moreover, in this model the basic "quantum cohomological"  
invariants are considered as reducable to integrals over the moduli  
space of instantons which is nothing more than the {\it quantizable}  
phase space of model. Thus, we are more or less concerning the  
quantum structure of the phase space of the model. Hence, the  
symplectic structure of the same moduli space/phase space is that  
structure which (after its quantization) allows one to define such  
quantum cohomological invariants.

In the second model in Ref. \cite{witva} the main structures used  
to identify the quantum cohomology (ring) of the theory are the  
existent Kaehler/Calabi-Yau structure and the $U(1)$ current of  
model. Recall that the calabi-Yau structure can be considered (in  
the quantization sheme) as equivalent to the (holomorphic)  
polarization of phase space where the symplectic form vanishes. It  
should be mentioned that the holomorphic polarization can be  
considered again as the flatness condition of the $U(1)$-connection  
of the quantization.

Thus, depending on the type of quantization we have various  
apparently independent conditions or geometrical structures on the  
phase space which are synonyms of each others in a more fundamental  
theory of quantization.

Thus, properly here also it is only the abstract quantum structure  
of the quantized phase space of the theory which determine the  
existence and the general abstract structure of the related quantum  
cohomology. Therefore, every quantized theory should have its own  
type of quantum cohomology, however the existence and abstract  
structure of these  quantum cohomologies should depend only on the  
general quantization structure, i.e. the line bundle/$U(1)$ or the  
Kaehler/Clabi-Yau structure of the related phase spaces.

To begin let us mention some usefull results on the classical  
cohomology that according to the selebrated de Rham's theorem one  
has $H^r \cong H_r$ and by the Hodge's theorem on a compact  
orientable Riemannian manifold we have $H^r (M) \cong Harm^r (M)$,  
where the $Harm^r$ is defined by $\Delta Harm^r = 0$. Moreover, we  
have according to the Poincare duality on a compact $m$-dimensional  
maifold $H^m \cong H^0$. On the other hand, we know from the most  
simple non-relativistic quantum mechanics that for the ground state  
$\Delta |0> = E_0 |0>$. Thus, in view of the fact that always $E_0  
\propto \hbar$ and that the classical limit of quantum mechanics is  
related with $\hbar \rightarrow 0$ \cite{s,h}, one has  
${\{|0>_{\hbar \rightarrow 0}}\} \in Harm^0$.

Therefore, one has for example for $H^0$ cohomology the isomorphism  
$H^0 \cong Harm^0 \cong H^0_{Q, \hbar \rightarrow 0}$, where  
$H^{0}_Q = H_{\psi_0}$ is the cohomology of ground state. In this  
sence one should obtain the usual "classical" cohomology as the  
classical limit of quantum cohomology, i. e. $H^m \cong Harm^m \cong  
H^0_{Q, \hbar \rightarrow 0}$ or one may consider the ground state  
as a quantum deformation of harmonic functions. Nevertheless, one  
can also consider that the quantum Laplace operator is a deformation  
of the usual Laplacian $\Delta_Q := \Delta + O(\hbar)$, whereas the  
ground state remains a harmonic function. Thus, one has at any case  
according to the ground state equation $\Delta |0> = E_0 |0>$ a  
deformed cohomology ring structure in the quantum case by the  
Hodge's theorem.

Now let us consider for the spatial base manifold of our quantum  
theory a Riemann surface $\Sigma$ with boundary, e. g. a disc. This  
is for example the case if we use to quantize a classical  
Chern-Simons-theory for QHE which is defined on $\Sigma \times R$  
\cite{meinq}. Despite of classical case where the "classical"  
boundary of $\partial \Sigma = C_1$ is well defined and we should  
have a classical mechanical prescription to determine such an  
"absolute" boundary. In quantized cases, i. e. if we have a quantum  
theory on $\Sigma$, the notion of boundary of $\Sigma_Q$ loses its  
definition and we have no quantum theoretical prescription to define  
or to measure the boundary $\partial \Sigma_Q$. In other words, if  
$\Sigma_Q$ is the polarized/quantized phase space, then in view of  
the uncertainty relation $\delta p \cdot \delta q = \hbar$ we have  
always $\delta q > 0$ and so it is $\partial \Sigma_Q \neq C_1$  
\cite{woodco}. Recall that the notion of polarization becomes  
equivalent to that of holomorphicity in the suitable almost complex  
or Kaehler cases, where the notion of J-holomorphic curves inters  
the quantum cohomology.

Therefore, the $\Sigma_Q$ is in quantized cases boundaryless in the  
usual "classical" sense of boundary or it is $\partial \Sigma_Q =  
\emptyset$. We are faced with such situation in the QHE, where the  
edge currents which should be classically exactly on the boundary of  
samples, are defined to flow within a width of magnetic length  
$l_B$ on the boundary of sample. Furthermore, also in QHE there is a  
potential drop around the boundary in a width of $l^{-1}_B$ which  
should not exists in classical case \cite{witerk}. These effects can  
be understood if one takes the electrodynamical uncertainty into  
account \cite{mdrop} where we have $e \delta A_m \cdot \delta q =  
\hbar$ with $\delta q := l_B$ which is confirmed not only by the  
flux quantization but also by definition of the magnetic length  
itself. Thus, in QED cases , e. g. in QHE which is the best known  
quantum effect in two dimensions, either one has to determine a new  
prescription to define the boundary of $\Sigma_Q$ or one may use the  
known methodes to define it with the help of edge currents or  
potential drops. At any case as it is mentioned above in quantum  
cases there is no possibility to define an absolute {\it one  
dimensional} boundary like the classical boundary $C_1$ for  
$\Sigma_Q$.
The quantum measurements and all possible quantum prescriptions can  
determine only a "quantum" boundary for $\Sigma_Q$ which is a two  
dimensional ring of  width $\delta q = l_B$.

It should be mentioned that, of course one is able to define a  
classical boundary for $\Sigma$ with the help of a classical theory,  
however this prescription and such a boundary are exact only within  
the classical limit (also of a quantum measurement) and the whole  
system is purely {\it classical}.

Now we use the above mentioned circumstances to show first the  
existence of a non-trivial cohomology in quantum case which becomes  
trivial in the classical case and second to prove its isomorphism  
with a related Floer cohomology. By the Floer cohomology we mean  
here the general cohomology of the ground state of the quantum  
system under concideration. If it is so, then in view of the already  
proved isomorphism between the Floer and quantum cohomology  
\cite{sadov} one should consider the mentioned non-trivial  
cohomology as a candidate of quantum cohomology.

Roughly speaking the quantum cohomology $H^r_Q$ is given by $H^r_Q  
:= H^r + \hbox{(additional terms)}$, nevertheless in view of the  
Hodge's theorem and the definition of $Harm^r$ one can use  
alternatively the deformation of the Laplace operator to define the  
quantum cohomology. We show that there is a non-trivial maximal  
cohomology $H^2_Q$ on $\Sigma_Q$ which is isomorphic to a $H^0_Q$ or  
to $Harm^0$, which is a $U(1)$-Floer cohomology.

Therefore, let us first define the above mentioned general Floer  
cohomology $ H_F$ according to the cohomology of ground state, i. e.

\begin{equation}
H_F := {\{ |0>, \qquad \Delta |0> = E_0 |0> }\} \qquad ; \qquad E_0  
\propto \hbar
\end{equation}
\label{one}

It is obviously a deformation of the cohomology of harmonic  
functions which is isomorphic to $H^0$ and in our case also it is  
also isomorphic to the $H^2$. Furthermore, it results in a  
deformation of the Laplace operator

\begin{equation}
\Delta_Q := \Delta +  O(\hbar) = \Delta - E_0 ,
\end{equation}
\label{two}

in the sense that now we have

\begin{equation}
\Delta_Q |0> = 0
\end{equation}
\label{twot}

It means that the deformed Laplacian has again the harmonic  
functions as its eigen vectors

$\Delta_Q Harm^0 = 0$ or $|0> \in Harm^0$. Furthermore, it requires  
also a deformations of the exterior differential operator and its  
adjoint:

\begin{equation}
d_Q := d + O(\hbar^{\frac{1}{2}}) \;, \; d_Q ^{\dagger} :=  
d^{\dagger} + O(\hbar^{\frac{1}{2}})
\end{equation}
\label{three}

This deformation requires also a deformation of the cup product  
which is essential in the usual definition of quantum cohomology and  
it can result also in a deformed differential structure of the  
quantum plane- or quantum group types \cite{next}.

More importent is the fact that, in view of the above analysis of  
quantum situation with $\partial \Sigma_Q = \emptyset$ we have a  
non-trivial maximal cohomology on $\Sigma_Q$ which is given  
according to the trivially closed but non-exact electromagnetic  
$2-form$:

\begin{equation}
H_Q^2 (\Sigma_Q ; F ) = {\{ F; dF = 0 ; F\neq dA }\} := {\{ dF = 0/  
F = dA}\}
\end{equation}
\label{four}

Obviously, such a cohomology can be defined also for any relevant  
general two form, i. e. for $\Omega^2$, on $\Sigma_Q$  instead of  
$F$.

Recall that this cohomology is trivial in the classical case where  
$\delta q = 0$ and we have $\partial \Sigma = C_1$. In this case as  
it is well known the closed $2-form F$ can always be considered as  
$F = dA$ or $\Omega^2 = d \Omega^1$. Thus, we have a classically  
trivial cohomology which become only in quantum case non-trivial.  
Recall further that according to the Hodge's decomposition in its  
$2-D$ case $\Omega^2 = d\Omega^1 \bigoplus Harm^2$ which applies  
also for our case, the closed form $F$ should be written as $F = dA  
\bigoplus Harm^2$.

Moreover, a deformation like (2) recalls one on the Witten's  
supersymmetric modification of Laplacian \cite{AtiWit} which should  
be discussed later.

Of course to show that $H^2_Q$ is the same as the known quantum  
cohomology \cite{witva} one should prove for example that the space  
of $2-forms {\{F}\}$ on the $U(1)$ bundle is the same as the space  
of J-holomorphic maps between Riemann surface and the base manifold  
of the discussed quantum theory \cite{manin}. This will be in our  
terminology a map between the Riamann surface and the polarized  
phase space of the quantum theory. We will prove this in a  
subsequent paper \cite{next}, however let us mention that in our  
case of $U(1)$ bundle we have a map from Riemann surface to the  
$2-D$ phase space or the moduli space of the  $U(1)$-connections  
which is again a Riemann surface in view of its Kaehler structure.

Moreover, as it is mentioned we have in view of the Hodge's theorem

\begin{equation}
H^2_Q \cong Harm^2
\end{equation}
\label{five}

and according to the Poincare duality

\begin{equation}
H^2_Q \cong H^0_Q
\end{equation}
\label{six}

Thus, we obtain the desired isomorphism between our quantum  
mechanically non-trivial cohomology $H^2_Q$ and the Floer cohomology  
of ground state $|0> \in Harm^0$ by the use of Hodge's theorem for  
$H^0 \cong Harm^0$:

\begin{equation}
H^2_Q \cong Harm^0 \cong H_F
\end{equation}
\label{seven}

As a conclussion we like to mention that the quantum theoretical  
version of the above mathematical prove of these equivalencies  
should be demonstrated as follows:

*) $H^2_Q$ is a result of $\delta q>0$ in quantum theory.

*) $\Delta_Q Harm^0 = E_0 Harm^0$ or $\Delta_Q |0> = E_0 |0>$ is a  
result of $ \delta E = 2E_0$ in quantum theory.

*) $H^2_Q \cong Harm^0 \cong H_F$ isomorphism is a result of the  
uncertainty relation:
$\delta q \cdot \delta p \cong \delta E \cdot \delta t = \hbar$.

\bigskip
Footnotes and references


\begin{thebibliography}{100}
\bibitem{vier}
We mean here the mechanics as the general abstract pattern of  
physical theories. By the global character of quantum observables we  
mean for example their spectral or integral representation which is  
related with the global character of wave function.
\end{thebibliography}
\end{document}